\documentclass[lettersize,journal]{IEEEtran}
\IEEEoverridecommandlockouts
\usepackage{cite}
\usepackage{amsmath,amssymb,amsfonts}
\usepackage{algorithm,algorithmic}	
\usepackage{graphicx}
\usepackage{textcomp}
\usepackage{mathtools}
\usepackage{tablefootnote}

\usepackage{multirow}
\usepackage{multicol}
\usepackage{soul,xcolor}

\usepackage{subcaption}
\usepackage[normalem]{ulem}

\newcommand{\boldmaxoper}{\mbox{\boldmath{Maximize:}}}
\newcommand{\boldvars}{\mbox{\boldmath{$(x,v,p^{S},p^{A})$}}}
\DeclareMathOperator*{\argmax}{arg\,max}
\DeclareMathOperator*{\argmin}{arg\,min}
\newcommand{\task}{\langle j, t \rangle}
\newcommand{\taskset}{\langle j, t \rangle \in \langle \mathcal{J}, \mathcal{T} \rangle}
\graphicspath{{figures/}} 

\begin{document}

\title{IoT-Aerial Base Station Task Offloading with Risk-Sensitive Reinforcement Learning for Smart Agriculture}
\author{Turgay Pamuklu$\dag$, \IEEEmembership{Member, IEEE}, \IEEEauthorblockN{Anne Catherine Nguyen$\dag$,  Aisha Syed$\ddag$, \\ W. Sean Kennedy$\ddag$, Melike Erol-Kantarci$\dag$, \IEEEmembership{Senior Member, IEEE}}

\IEEEauthorblockA{$\dag$\textit{School of Electrical Engineering and Computer Science,}
\textit{University of Ottawa}, Ottawa, Canada}

\IEEEauthorblockA{$\ddag$\textit{Artificial Intelligence Research Lab, Nokia Bell Labs, Murray Hill, NJ, USA}\\
Emails:\{turgay.pamuklu, anguy087, melike.erolkantarci\}@uottawa.ca, \\
\{aisha.syed, william.kennedy\}@nokia-bell-labs.com}
}
\maketitle
\makeatletter
\def\ps@IEEEtitlepagestyle{%
  \def\@oddfoot{\mycopyrightnotice}%
  \def\@oddhead{\hbox{}\@IEEEheaderstyle\leftmark\hfil\thepage}\relax
  \def\@evenhead{\@IEEEheaderstyle\thepage\hfil\leftmark\hbox{}}\relax
  \def\@evenfoot{}%
}
\def\mycopyrightnotice{%
  \begin{minipage}{\textwidth}
  \centering \scriptsize
Accepted Paper. DOI:10.1109/TGCN.2022.3205330. IEEE policy provides that authors are free to follow funder public access mandates to post accepted articles in repositories. When posting in a repository, the IEEE embargo period is 24 months. However, IEEE recognizes that posting requirements and embargo periods vary by funder. IEEE authors may comply with requirements to deposit their accepted manuscripts in a repository per funder requirements where the embargo is less than 24 months.
  \end{minipage}
}
\makeatother

\begin{abstract}
Aerial base stations (ABSs) allow smart farms to offload processing responsibility of complex tasks from internet of things (IoT) devices to ABSs. IoT devices have limited energy and computing resources, thus it is required to provide an advanced solution for a system that requires the support of ABSs. This paper introduces a novel multi-actor-based risk-sensitive reinforcement learning approach for ABS task scheduling for smart agriculture. The problem is defined as task offloading with a strict condition on completing the IoT tasks before their deadlines. Moreover, the algorithm must also consider the limited energy capacity of the ABSs. The results show that our proposed approach outperforms several heuristics and the classic Q-Learning approach. Furthermore, we provide a mixed integer linear programming solution to determine a lower bound on the performance, and clarify the gap between our risk-sensitive solution and the optimal solution, as well. The comparison proves our extensive simulation results demonstrate that our method is a promising approach for providing a guaranteed task processing services for the IoT tasks in a smart farm, while increasing the hovering time of the ABSs in this farm.    
\end{abstract}
\begin{IEEEkeywords}
Energy Efficient Computation Offload, Unmanned aerial vehicle (UAV), Aerial base station (ABS), Smart farm, Risk-Sensitive Reinforcement Learning, Internet of Things (IoT).
\end{IEEEkeywords}
\section{Introduction}
\par Agriculture is sensitive to environmental changes and even slight changes can alter the outcome of the crops or livestock. Smart farms use Internet of Things (IoT) sensors, such as cameras, to monitor such environmental changes. The sensors can capture images to monitor the status of hundreds of acres of land. We can then use the captured images and image recognition techniques to extract valuable information about the status of the farm such as, the presence of fire, the presence of pests, and the growth rate of the crops and livestock. Jhuria et al. trained neural networks to recognize certain diseases on crops during the growing process \cite{Jhuria2013}. Dhumale et al. introduced a robot that captures images of crops, and use image processing techniques to detect pests and diseases on the crops, and spray the appropriate crops with pesticide \cite{Dhumale2021}. IoT cameras have become essential wireless sensors for smart agriculture and precision farming because they provide real-time information on the status of the farm and they relay that information to a central location. With IoT sensors, farmers now have the capability to monitor multiple acres of land from one location and provide precise care to their farm. 
\par Image processing is an intensive task and the central processing units on the IoT sensors may not have the capacity to perform such intensive tasks. In addition, some image processing tasks can be time-sensitive such as fire detection. In such circumstances, we can make use of aerial base stations (ABS) that can connect the sensors to much more powerful computational resources. The ABSs equipped with computational units can perform the tasks themselves, or relay the tasks to a nearby multi-access edge computing (MEC) device. This enables the task to be done before the deadline.

\par Because the ABSs are hovering above large remote farmlands, they must be battery operated. As previously mentioned, image processing tasks are computationaly intensive and will require large amounts of power. If the ABS does too many tasks, it can shorten the longevity of the ABS' operation time. On the other hand, without the ABS, the IoT sensors cannot send their tasks and the sensor network is no longer operational. Therefore, we aim to extend the longevity of the battery life of the ABSs. However, if the ABS does not perform any tasks in order to conserve its energy, the MEC device would have to do all of the tasks, then the tasks may not be able to meet their deadline. This may lead to detrimental consequences such as significant loss of farmland due to a slow fire detection. Therefore the network needs to have a decision making process that considers both the time-critical nature of the image processing tasks, as well as controlling the energy consumption of the ABSs in order to maximize their longevity.  

\par Every decision comes with tradeoffs, especially in a system with multiple objectives. The network's decision making algorithm must decide which resources will process the task based on several factors such as the ABS' current energy level, and the deadline of the task. It must also prioritize between the two objectives, the ABS network's longevity, or the potential costs of not meeting the deadline. But, missing the image processing tasks' deadlines potentially can have damaging consequences, therefore the decision making process must also evaluate the risk of each possible outcome and select the optimal decisions that will minimize the risk.  

\begin{figure*}
\centering
\includegraphics[width=0.98\textwidth]{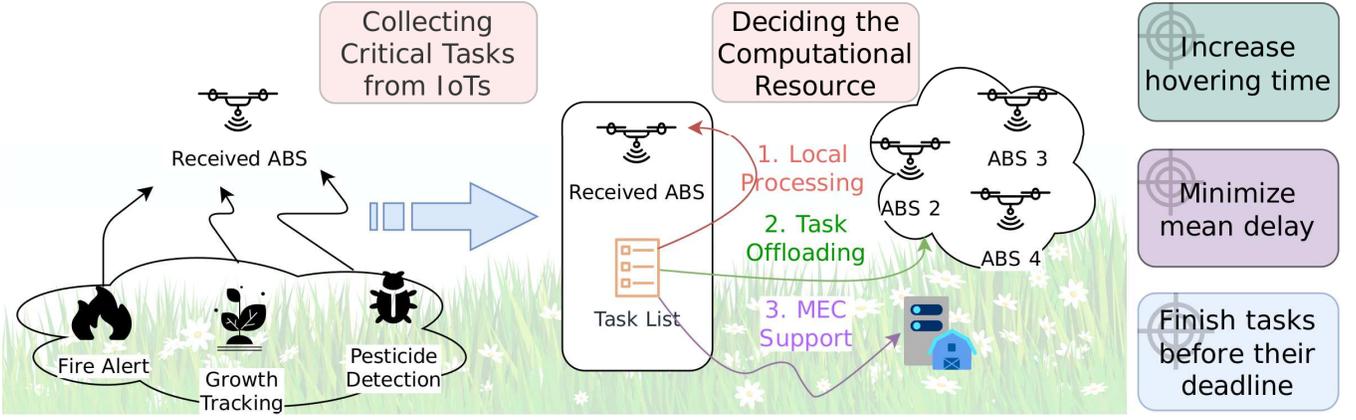}
\caption{\label{fig:arch} Illustration of the proposed smart farm.}
\end{figure*}
\subsection{Main Contributions}
\par Machine learning is a promising approach for solving wireless network optimization problems \cite{Akbari2021,Khoramnejad2021,Pamuklu2021}. In our previous study, we presented a Q-Learning solution for a joint task-oriented delay-aware offloading as well as increased ABS hovering time \cite{Nguyen2021}. Following our prior work, in this paper, we propose a risk-sensitive reinforcement learning (RL) approach to guarantee that deadlines will be met for IoT devices in our smart farm network. By using this method, we isolate the constraints from the other KPIs, and dynamically change the weight of the KPIs while training the model.  The results reported in this paper prove that our risk-sensitive method significantly reduces the number of deadline violations and constrains it to a specific threshold. To the best of our knowledge, the multi-agent-based risk-sensitive RL method is proposed for the first time for a constrained MDP problem in this paper. Our contributions can be summarized as follows:
\begin{enumerate}
\item We solved a constrained MDP problem by modifying the training phase of a machine learning (ML) algorithm for an aerial base station (ABS)-supported smart farm environment. We created a Q-table that is trained separately from the other objectives in the problem. Therefore, we could have determined the borders of a risky task-oriented constraint and adaptively changed the direction of the learning process by considering the threshold of this constraint.
\item We implemented this method with a multi-agent RL approach. Thus, this method can address larger networks due to the scalability capability of this approach for this risk-sensitive-based problem. 
\item We provided an upper-bound analysis with a MILP Solver. Therefore, we evaluated our RL approach with the different weights for our multi-objective NP-Hard problem. 
\item We accomplished several stress tests on the proposed solution. Thus, we analyzed the robustness of the proposed method in cases of unexpected changes in the smart farm environments.
\end{enumerate}

The rest of the article is organized as follows. The related work is discussed in Section II. Then, in Section III we describe our smart farm model and its problem definition. Section IV details the risk-sensitive RL approach, and Section V provides the detailed results of this approach. Finally, in Section VI, we conclude the paper.

\section{Related Works}
\par In one of the recent unmanned aerial vehicle (UAV) based studies, Liao et al. decomposed their task offloading and resource allocation problems into two subproblems \cite{Liao2021}. Then, they solved these two problems separately with an actor-critic-based learning algorithm and a heuristic algorithm. As a result, they demonstrated that their proposed method reduces queuing delay more than their referenced solutions. A similar problem is addressed in \cite{Xiong2022} using the successive convex approximation method. The task-oriented guaranteed delay is a main objective in our paper, and Zhao et al. also focused on this guaranteed delay problem \cite{Zhao2020}. However, they preferred the relaxation techniques to solve this hard-constrained problem. Khairy et al. also used Lagrangian relaxation for the constraints in their UAV-based communication problem \cite{Khairy2020}. Next, they solved this dual problem with a deep RL algorithm. Ghdiri et al. have a distinctive study that also targets the task deadline and the limited UAV battery problems \cite{Ghdiri2020}. However, their aim was to find the optimum cluster heads' locations and UAV trajectory planning. Zhang et al. also aimed at the trajectory planning problem and addressed it by proposing a Safe-DQN solution \cite{Zhang2021}.

\par Yang et al. minimized the total energy consumption of UEs and UAVs in their wireless network \cite{Yang2019}. They have proposed several heuristics for different subproblems of their main nonconvex problem. Another heuristic solution for a UAV network was suggested by Xu et al., where the authors optimized a broad number of critical decisions \cite{Xu2021}. As an alternative for heuristic approaches, Yao et al. introduced a game-theoretic solution for resource allocation and the channel access problem \cite{Yao2020}. They provided a promising method in terms of reducing the total energy consumption in their proposed network. Another energy-efficient solution is proposed in \cite{Wang2021}, where UAVs process and relay cell-edge users' tasks.

\par Risk-sensitive RL has attracted attention in recent wireless networks papers due to its promising performance to solve constrained problems. Khalifa et al. used this method for targeting strict latency requirements of URLLC traffic \cite{Khalifa2019}. Their results showed that they can increase the number of successfully received URLLC packets while keeping the number of lost packets lower than a threshold value. The example of another wireless network solution was suggested by Alsenwi et al., where they used their novel "conditional value at risk" approach to prevent eMBB users from getting lower data rates due to concurrent URLLC traffic in the network \cite{Alsenwi2019}.

\par In their task offloading problem, Zhou et al. extended the available processing resources by adding satellites to their network \cite{Zhou2021}. Then, they proposed a centralized deep risk-sensitive solution to achieve their energy consumption constraint. Despite the fact that this paper has a related problem, we provide a task-oriented approach to finish the IoT tasks before their deadline. Moreover, our risk-sensitive solution is a distributed approach in which each ABS has its own intelligent local agent to make offloading decisions. To the best of our knowledge, this is the first work that proposes a multi-agent-based risk-sensitive RL solution for the task offloading problem in a UAV/ABS based wireless network.

\begin{table}
\centering
\caption{\label{tab:Notations} Summary of the notations.}
\begin{tabular}{c|c|p{4.6cm}}
\textbf{Sets} & \textbf{Size} & \textbf{Description} \\ \hline
$t\in\mathcal{T}$ & $T$ & set of time intervals \\
$j\in\mathcal{J}$ & $J$ & set of ABSs  \\
$\taskset$ & $J, T$ & tuple of tasks  \\
$l\in\mathcal{L}$ & $L$ & set of MEC devices \\
$j'\in\mathcal{J}^{+}$ & $J$+$L$ &set of computational resources \\
$k\in\mathcal{K}$ & $K$ & set of task types  \\ \hline
\textbf{Task Param.}& \textbf{Range} & \textbf{Description} \\ \hline
$\alpha^{B}_{\task }$ & $\{0,1\}$ & a task comes to ABS j in time interval t \\
$\alpha^{P}_{\task }$ & $\mathbb{R}$ & the processing time of this task\\
$\alpha^{I}_{\task }$ & $\mathbb{R}$ & the transmission delay between IoT $\&$ ABS for this task\\
$\alpha^{D}_{\task }$ & $\mathbb{R}$ & the deadline for this task\\ 
$\Delta_{\task }$ & $\mathbb{R}$ & end-to-end delay of this task \\ \hline
\textbf{Energy Param.}& \textbf{Range} & \textbf{Description} \\ \hline
$\Upsilon^{B}_{j}$ & $\mathbb{N}$ & battery capacity in ABS $j$ \\
$\Upsilon^{H}_{j}$ & $\mathbb{R}$ & hovering energy cons. in ABS $j$ \\	
$\Upsilon^{A}_{j}$ & $\mathbb{R}$ & signal trans. energy cons. in ABS $j$\\	
$\Upsilon^{I}_{j}$ & $\mathbb{R}$ & idle energy cons. in ABS $j$\\
$\Upsilon^{C}_{j}$ & $\mathbb{R}$ & computing energy cons. in ABS $j$\\
$\Upsilon^{R}_{j}$ & $\mathbb{R}$ & remaining energy in ABS $j$\\ \hline
\textbf{Weight Param.}& \textbf{Range} & \textbf{Description} \\ \hline
$\Upsilon^{L}_{j}$ & $\mathbb{N}^{+}$ & battery reward level\\
$\mathcal{V}^{L}_{j}$ & $\mathbb{N}$ & violation reward level\\ 
$W$ & $[0,1]$ & energy consumption weight\\
$\Theta^{D}$ & $\mathbb{R}$ & scaling factor for deadline violation\\
$\Theta^{M}$ & $\mathbb{R}$ & scaling factor for mean delay\\ \hline
\textbf{Variables}& \textbf{Domain} & \textbf{Description} \\ \hline
$x_{\task j'}$ & $\{0,1\}$ & task $\task$ is processed in resource $j'$ \\
$p^{A}_{\task j't'}$ & $\{0,1\}$ & the time intervals ($t'$) used to processed the corresponding task\\
$p^{S}_{\task j't'}$ & $\{0,1\}$ & the starting time interval ($t'$) for the corresponding task\\
$p^{E}_{\task j't'}$ & $\{0,1\}$ & the ending time interval ($t'$) for the corresponding task\\
$v_{\task }$ & $\{0,1\}$ & the corresponding task does not finish before its deadline\\ 
\hline
\end{tabular}
\end{table}

\section{System Model and Problem Formulation}
\par In this section, we briefly describe the scenario, detail the energy consumption and delay models, specify the system constraints, and finally formulate the guaranteed deadline problem \cite{Nguyen2021}. Table~\ref{tab:Notations} summarizes the notations we use in the remainder of the article.
\subsection{Scenario}
\par Figure~\ref{fig:arch} illustrates the scenario we use in this study. We assume a time-interval-based model. In each time interval ($t\in\mathcal{T}$), a set of ABS ($j\in\mathcal{J}$) devices collect the tasks from IoT devices scattered accross a smart farm. If ABS $j$ receives a task in time interval $t$, the decision-maker in this ABS is informed by the $\alpha^{B}_{\task }=1$ indicator \footnote{The length of a time interval is significantly small. Therefore, an ABS receives at most one task in a time interval. Hence, we may represent a task with a tuple ``$\task$" in which $j$ is the received ABS and $t$ is the received time interval.}. Then, ABS $j$ decides on a computational resource ($j'\in\mathcal{J}^{+}$) to process the task while taking into account the type of task ($k\in\mathcal{K}$), and the network conditions such as the expected transmission delay to offload a task to another ABS. Furthermore, the computational resource set is not limited to just ABS devices, but also contains nearby MEC devices ($l\in\mathcal{L}$).
\par \textbf{Given Data:} 
\begin{itemize}
\item A binary indicator signifying that ABS $j$ received a task in time interval $t$ ($\alpha^{B}_{\task }$). The task's processing time ($\alpha^{P}_{\task }$) and deadline ($\alpha^{D}_{\task }$). 
\item Transmission delays between IoTs and ABSs ($\alpha^{I}_{\task }$). 
\item Energy consumption of each component in the ABSs ($\Upsilon^{H}_{j}, \Upsilon^{A}_{j}, \Upsilon^{I}_{j}, \Upsilon^{C}_{j}$). 
\item Battery capacity of each ABS ($\Upsilon^{B}_{j}$).
\end{itemize}
\par \textbf{Objectives:}
\begin{itemize}
\item Increasing the hovering time of ABSs. We aim to maximize the operating time before we need to recharge any of the ABSs in our smart farm. 
\item Reducing the mean end-to-end delay for processing of the tasks generated by IoTs.
\item Ensure that the number of deadline violations will not exceed a certain amount.
\end{itemize}
\par \textbf{Decisions:}
\begin{itemize}
\item Each ABS $j$ should independently decide the optimum resource to process the received task $\langle j,t \rangle$. If an ABS chooses to process that task locally, it will queue the request and begin to process it whenever its processing resource is available. Otherwise, the ABS will offload the task to another ABS or a MEC device in the smart farm.
\end{itemize}
\subsection{Energy Consumption and Delay Models}
\par In the interest of increasing the hovering time, we have to find the change in the remaining battery energy for each ABS ($\Upsilon^{R}_{j'}$) according to their decision. Eq. ~\ref{eq:encalc} calculates the remaining battery level where $\Upsilon^{H}_{j'}$, $\Upsilon^{A}_{j'}$, $\Upsilon^{I}_{j'}$ are the hovering, signal transmission, and idle energy consumptions respectively. We assume these values do not change with the offloading decision, and they are fixed values during our evaluation period ($T$). On the other hand, computing energy consumption, $\Upsilon^{C}_{j'}$, is determined by task processing decisions ($p^{A}_{\task j't'}$) which equals to one if the resource processes a task.       
\begin{flalign}
\label{eq:encalc}
\Upsilon^{R}_{j'} = & \Upsilon^{B}_{j'} - (\Upsilon^{H}_{j'} + \Upsilon^{A}_{j'} + \Upsilon^{I}_{j'}) * T  \notag \\ &-\sum\limits_{t'\in\mathcal{T}}\sum\limits_{\taskset} (\Upsilon^{C}_{j'} - \Upsilon^{I}_{j'}) * p^{A}_{\task j't'}&
\end{flalign}

\par Finalizing each task before their deadline is another one of our key performance indicators (KPI). Therefore, for each task, we have to calculate the time difference between the task's generation at the IoT device and the task's completion at the computational resource. Eq. ~\ref{eq:delay_per_task} formulates this in which $(t')$ is the start time of the task at the resource. That starting time is determined by the binary decision variable $p^{S}_{\task j't'}$, which equals to one if resource $j'$ has started to process that task at time interval $t'$. At any other time, it is equal to zero; thus, the summation of the first term equals $t'$. The other terms in this equation are $t$, $\alpha^{I}_{\task }$, and $\alpha^{P}_{\task }$, which are the received time of the task at ABS, the transmission delay between IoT and ABS, and the processing delay, respectively. Lastly, we calculate the mean delay with Eq. ~\ref{eq:mean_delay}, in which we sum the end-to-end delays of all tasks and then divide it by the total number of tasks.     
\begin{flalign}
	\label{eq:delay_per_task}
	& \Delta^{E}_{\task } = \sum\limits_{j'\in\mathcal{J}^{+}} \sum\limits_{t'\in\mathcal{T}} p^{S}_{\task j't'} * (t') - t  + \alpha^{I}_{\task } + \alpha^{P}_{\task }  \\ 
	\label{eq:mean_delay}
	& \delta = \frac{\sum\limits_{\taskset} \Delta^{E}_{\task } * \alpha^{B}_{\task }}{\sum\limits_{\taskset}\alpha^{B}_{\task }} 
\end{flalign}

\par In Eq.~\ref{eq:delayviol1}, $\alpha^{D}_{\task }$, $\Delta^{E}_{\task }$, and $v_{\task }$ are the deadline, end-to-end delay, and violation indicator of task $\task$, respectively. In addition, we have an $M$ term which is a significantly large number. In the case of the deadline violation ($\alpha^{D}_{\task } - \Delta^{E}_{\task } < 0$), the left side of this equation will be lower than zero. Thus, to grant Eq.~\ref{eq:delayviol1}, the right side of this equation is enforced to be lower than zero, which could be achieved by only $v_{\task }=1$. Contrastingly, if no deadline violation has occurred, the left side of this equation will be higher than zero  ($\alpha^{D}_{\task } - \Delta^{E}_{\task } \geq 0$). Hence, $v_{\task }$ may equal either to zero or one. Therefore, Eq.~\ref{eq:delayviol2} will model that case. In this equation, if we can finish a task before its deadline ($\alpha^{D}_{\task } - \Delta^{E}_{\task } \geq 0$), the left side will be higher than zero. Thus, the right side of this equation should also be higher than zero, which can only be realized by $v_{\task }=0$.    
\begin{flalign}
	\label{eq:delayviol1}
	&\alpha^{D}_{\task } - \Delta^{E}_{\task } \geq  - M * v_{\task }, \quad\quad\quad\forall \taskset \\
	\label{eq:delayviol2}
	&\alpha^{D}_{\task } - \Delta^{E}_{\task } < M * (1-v_{\task }), \;\quad \forall \taskset
\end{flalign}

\subsection{System Constraints}
\par This subsection presents the constraints of our optimization problem. Eq.~\ref{eq:pulotlimit} denotes that a computational resource ($j'\in\mathcal{J}^{+}$) should process at most one task in a time interval ($t'\in\mathcal{T}$). Eq.~\ref{eq:ABSlimit} guarantees that a task is processed by at most one resource in a time interval ($t'\in\mathcal{T}$). 
\begin{flalign}
    \label{eq:pulotlimit}
	&\sum\limits_{\taskset} p^{\varpi}_{\task j't'}  \leq 1, \notag \\
	& \qquad\qquad\forall j'\in\mathcal{J}^{+} , \forall t'\in\mathcal{T}, \forall \varpi \in \{A, S, E\} \\
	\label{eq:ABSlimit}
	&\sum\limits_{j' \in \mathcal{J}^{+}}p^{\varpi}_{\task j't'}  \leq 1, \notag \\	
	&\qquad\qquad \forall \taskset, \forall t'\in\mathcal{T}, \forall \varpi \in \{A, S, E\}
\end{flalign}	
\par Eq.~(\ref{eq:contiguity}-\ref{eq:contiguity4}) maintain the association between the task processing binary decision indicators. Assume that the resource $j'$ is idle in time interval $t'$, which yields $p^{A}_{\task j't'} = 0$. Then, it starts to process task $\task$ in the next time interval ($t'+1$), which leads to $p^{A}_{\task j'(t'+1)} = 1$. Hence, the starting time indicator in this time interval ($p^{S}_{\task j'(t'+1)}$) also becomes one due to Eq.~\ref{eq:contiguity}. In addition, the ending time indicator in this time interval ($p^{E}_{\task j'(t'+1)}$) equals zero due to Eq.~\ref{eq:contiguity2}.
\begin{flalign}
	\label{eq:contiguity}
	&p^{A}_{\task j'(t'+1)} = p^{A}_{\task j't'}+p^{S}_{\task j'(t'+1)} - p^{E}_{\task j'(t'+1)} \\
	\label{eq:contiguity2}
	&p^{S}_{\task j'(t'+1)} + p^{E}_{\task j'(t'+1)} \leq 1,    \notag \\
	&\qquad\qquad\forall \taskset, \forall j'\in\mathcal{J}^{+}, \forall t' \in \mathcal{T}  \\
    \label{eq:contiguity3}
	&p^{A}_{\task j'(0)} = p^{S}_{\task j'(0)}  \\ 
	\label{eq:contiguity4}
	&\sum\limits_{\substack{t'\in\mathcal{T}}} p^{\varpi}_{\task j't'} \leq 1, \notag \\
	&\qquad\qquad\forall \taskset, \forall j'\in\mathcal{J}^{+}, \forall \varpi \in \{S, E\}
\end{flalign}	
\par In addition, for a special condition in which a computational resource starts to process a task at the beginning of the simulation ($t'=0$), we need Eq.~\ref{eq:contiguity3} to indicate the starting time ($p^{S}_{\task  j'(0)} = 1$). Now assume that in time interval $t'$, the $j'$ processes the task $\task$, which yields $p^{A}_{\task  j't'} = 1$. Then, it stops processing that task in the next time interval ($t'+1$), which leads to $p^{A}_{\task  j'(t'+1)} = 0$. Therefore, the ending time indicator turns to one ($p^{E}_{\task  j't'} = 1$) due to Eq.~\ref{eq:contiguity}. Lastly, Eq.~\ref{eq:contiguity4} ensures that a task has only one starting and ending time. 
\begin{flalign}
    \label{eq:CRalloc}
	&\sum\limits_{t'=t}^{T}\sum\limits_{j' \in \mathcal{J}^{+}} p^{A}_{\task  j't'} * x_{\task  j'} = \alpha^{B}_{\task  } * \alpha^{P}_{\task  }, \notag \\
	&\;\;\qquad\qquad\qquad\qquad\qquad\qquad\qquad\quad\forall\taskset \\
	\label{eq:CRnegalloc}
	&\sum\limits_{t'=0}^{t-1}\sum\limits_{j' \in \mathcal{J}^{+}} p^{A}_{\task  j't'} * x_{\task  j'} = 0 ,\quad\quad \forall\taskset \\
	\label{eq:offloadlimit}
	&\sum\limits_{j'\in\mathcal{J}^{+}} x_{\task j'}  \leq 1 ,\qquad\qquad\qquad\qquad \forall \taskset 
\end{flalign}
\par Each received task ($\alpha^{B}_{\task }=1$) must be processed by a resource for a time period equal to its processing time ($\alpha^{P}_{\task }$). Eq.~\ref{eq:CRalloc} maintains that if a task is offloaded to $j'$, the offloading decision equals one ($x_{\task j'}=1$), and that equation simplifies as $\sum\limits_{t'=t}^{T}p^{A}_{\task j't'} = \alpha^{P}_{\task }$. Thus, we have to allocate $\alpha^{P}_{\task }$ number of time intervals at $j'$. Eq.~\ref{eq:CRnegalloc} ensures that a computational resource starts to process a task after it has received the task, meaning that task processing time intervals ($t'\in T$) should be higher or equal to task received time intervals ($t\in T$). Lastly, Eq.~\ref{eq:offloadlimit} bounds the number of offloading decisions for each task to prevent a multiple task offloading case.

\subsection{Problem Formulation}
\par We have two problem definitions, P1 and P2, for the smart farm scenario. The main difference between these two definitions originates from their perspective on deadline violation. While minimizing the number of deadline violations is included in the objective function of P1, the approach on P2 is to limit the number of deadline violations to a threshold (Eq.~\ref{eq:deadlineconstraint}).  
\begin{flalign}
\textbf{(P1)} \underset{\boldvars}{\boldmaxoper}\quad 
&W * \min_{ j' \in J} \Upsilon^{R}_{j'} - \frac{1-W}{2*\Theta^{M}} \delta \notag \\
&- \frac{1-W}{2*\Theta^{D}} \sum\limits_{\taskset}v_{\task }&\label{eq:objp1}\\
\textbf{Subject to: }\quad&Eqs.~(\ref{eq:pulotlimit}-\ref{eq:offloadlimit})& \notag
\end{flalign}
\par Eq.~\ref{eq:objp1} depicts the objective function of the first problem, in which $W$, $\Theta^{M}$, and $\Theta^{D}$ are the weight of the hover time, mean delay, and deadline violation scaling factors, respectively. $\Upsilon^{R}_{j'}$ is the remaining battery energy of ABS $j'$, which is calculated by Eq.~\ref{eq:encalc}. Task processing decisions ($p^{A}_{\task}$) play a significant role on this calculation. $\delta$ is the mean delay of all received tasks, and $v_{\task}$ is the indicator of deadline violation of task $\task$. Mean delay is calculated by Eq.~\ref{eq:mean_delay} and the deadline violation is determined by Eqs.~(\ref{eq:delayviol1}-\ref{eq:delayviol2}). The starting time decision of each task $p^{S}_{\task}$ affects these equations. Lastly, these two decision variables ($p^{A}_{\task}$, $p^{S}_{\task}$) are limited by the corresponding task's offloading decision ($x_{\task}$), as shown in Eqs.~(\ref{eq:CRalloc}-\ref{eq:CRnegalloc}).
\par As we mentioned earlier, the deadline violation KPI is not included in the objective function of P2 (Eq.~\ref{eq:objp2}). Instead, it is considered as a constraint in Eq.~\ref{eq:deadlineconstraint}, in which $\mathbb{V}$ is the upper bound of the allowed number of deadline violations.   
\begin{flalign}
\textbf{(P2)} \underset{\boldvars}{\boldmaxoper}\quad
\label{eq:objp2}
&W * \min_{ j' \in J} \Upsilon^{R}_{j'} - \frac{1-W}{\Theta^{M}} \delta& \\
\textbf{Subject to:} \quad& 
\label{eq:deadlineconstraint}
\sum\limits_{\taskset}v_{\task } \leq \mathbb{V}&\\
& Eqs.~(\ref{eq:pulotlimit}-\ref{eq:offloadlimit}) \notag
\end{flalign}
\par Task offloading problems are generalized assignment problems that are NP-Hard \cite{Zhang2020}. Therefore, we solve these two problems with a MILP solver and we propose machine learning-based solutions. The latter is explained explicitly in the next section.

\begin{figure*}[t]
  \begin{subfigure}{.49\textwidth}
  \includegraphics[width=.9\textwidth]{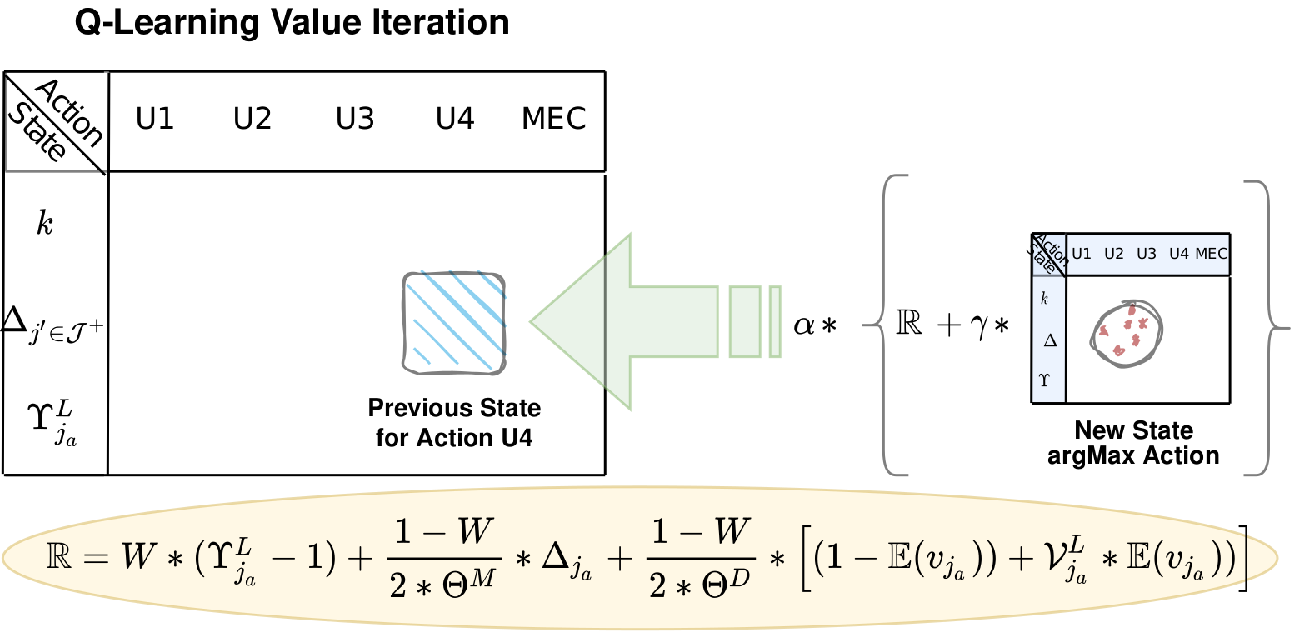}
  \caption{Value iteration for Q-Learning method.}\label{fig:valiter1}
  \end{subfigure}
  \begin{subfigure}{.49\textwidth}
  \includegraphics[width=.9\textwidth]{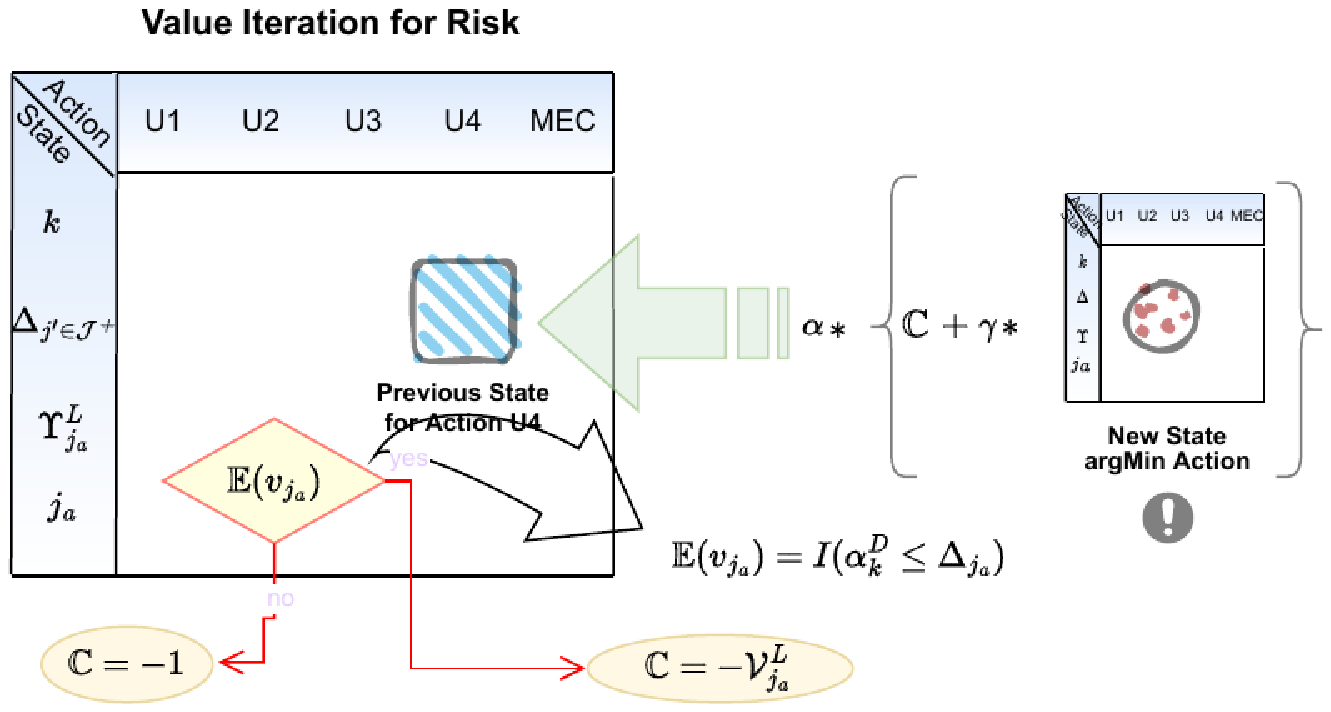}
  \caption{Value iteration for risk.}\label{fig:valiter2}
  \end{subfigure}
  \begin{subfigure}{.49\textwidth}
  \includegraphics[width=.9\textwidth]{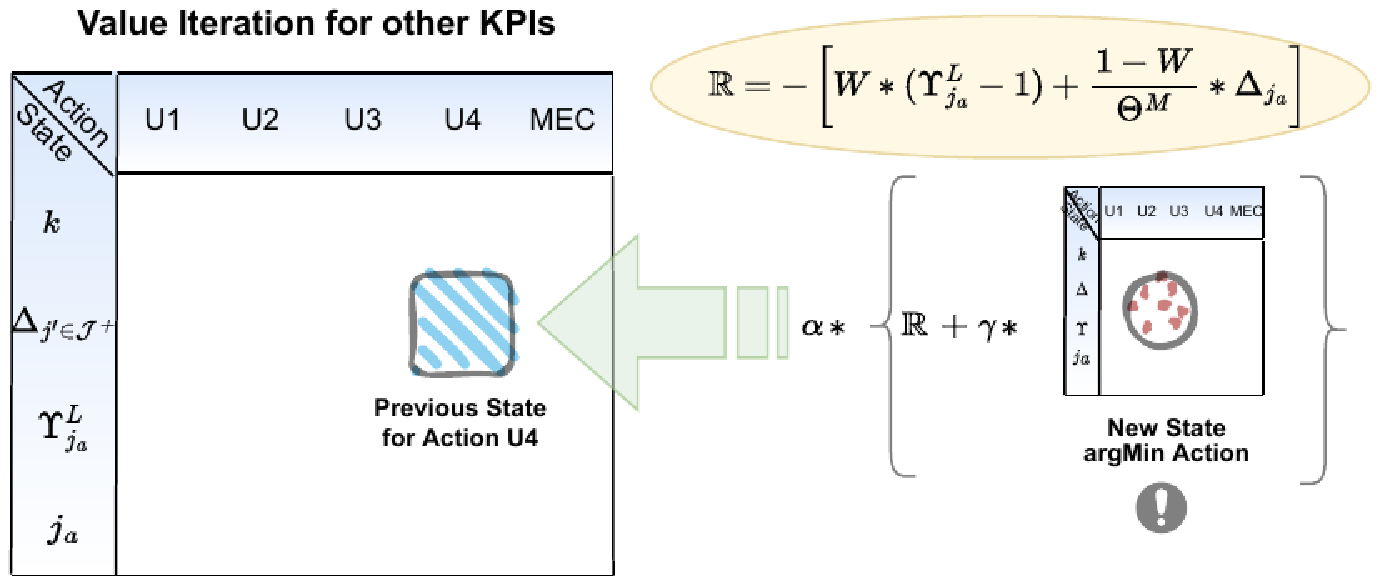}
  \caption{Value iteration for immediate reward.}\label{fig:valiter3}
  \end{subfigure}\hfill
  \begin{subfigure}{.49\textwidth}
  \includegraphics[width=0.9\textwidth]{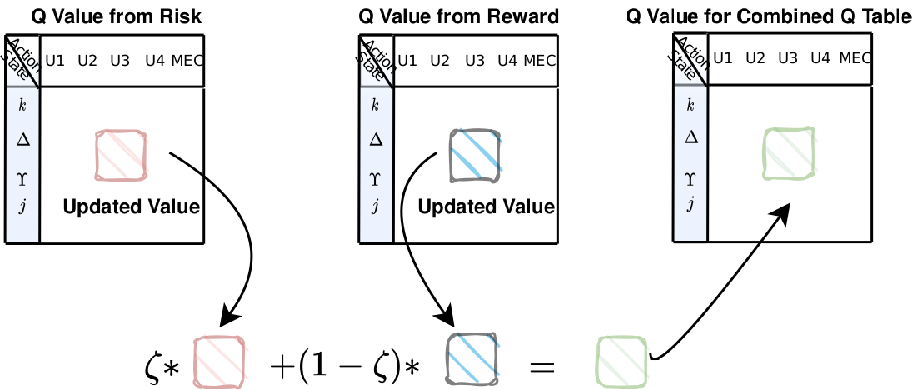}
  \caption{A third Q-table with a dynamically changed $\zeta$ value.}\label{fig:valiter4}
  \end{subfigure}
  \caption{Value iteration processes in Q-Learning and Risk-sensitive RL methods for an example scenario with one MEC server and 4 ABSs (U1 to U4).}
  \label{fig:valiter}
\end{figure*}

\section{Proposed Machine Learning-based Methods}
\par This section details the reinforcement learning algorithms we use to solve the explained smart farm problems. These algorithms run independently in each ABS in a distributed manner which improves the scalability of the proposed solutions. In the following subsection, we start to explain a solution for P1.
\subsection{Q-Learning Approach }
\label{sec:ql}
\par P1 (Eq.~\ref{eq:objp1}) has three KPIs in the objective function, and a finite-horizon MDP can address them with a tuple $\mathcal{F} = \{\mathbb{P}, \mathbb{A}, \mathbb{R}, \mathbb{S}, \Pi \}$. The details of that tuple are:
\begin{itemize}
    \item \textbf{State Transitions:} $\mathbb{P}:\mathbb{S}_{1}x\mathbb{A}x\mathbb{S}_{2}\implies \mathbb{R}$
    \par A transition is triggered whenever an ABS receives a new task:
    \begin{enumerate}
        \item First, the agent analyzes its current state $\mathbb{S}_{1}$.
        \item Second, it makes a decision for the action $\mathbb{A}$.
        \item Finally, it observes the environmental changes due to its action and determines the new state $\mathbb{S}_{2}$ and the immediate reward $\mathbb{R}$.
    \end{enumerate}
    Because of the uncertainty of task arrivals and the stochastic properties of the wireless channel model, the state transitions are not deterministic.     
    \item \textbf{Action:} $\mathbb{A}= \{j_{a} | j_{a} \in \mathcal{J^{+}}\}$
    \par Delegating the received task to a computational resource ($j_{a}$) is the action that an ABS agent should do in this MDP framework. The resource may be at the same ABS, another ABS, or a MEC. Therefore the action domain includes all of these options ($\mathcal{J}^{+}$).
    \item \textbf{State:} $\mathbb{S}= \{k, \{\Delta_{j'} | j'\in\mathcal{J^{+}} \}, \{\Upsilon^{L}_{j} | j\in\mathcal{J}$\}\}
    \par Task type ($k$) contains two pieces of information (task processing time and deadline) for determining the impact of an action on the system performance. Second, the expected delay ($\Delta_{j'}$) is crucial for calculating the system mean delay and deadline violations. Finally, ABS battery levels ($\Upsilon^{L}_{j'}$) should be included in the state space in order to improve the hover time KPI.   
    \item \textbf{Policy($\Pi$):} We choose the epsilon-greedy policy for the tradeoff between exploration and exploitation in the learning process. Also, we use the value iteration method to compute the optimal policy illustrated by Fig.~\ref{fig:valiter1}. In this figure, $\alpha$, $\gamma$, and $\mathbb{R}$ are the learning rate, discount, and immediate reward, respectively. The latter is detailed in the following paragraph. 
    \item \textbf{Reward:} 
    \begin{flalign}
    \label{eq:rew1}
    &\mathbb{R} = W * (\Upsilon^{L}_{j_{a}} - 1) - \frac{1-W}{2*\Theta^{M}} * \Delta_{j_{a}} \notag \\  &+\frac{1-W}{2*\Theta^{D}} *\left[ (1-\mathbb{E}(v_{j_{a}})) + \mathcal{V}^{L}_{j_{a}} * \mathbb{E}(v_{j_{a}})\right] \\
    \label{eq:rew2}
	& \Upsilon^{L}_{j_{a}} =
	\begin{cases} 
	        2, & \text{if } \mathbb{E}(\Upsilon^{R}_{j_{a}}) - \max_{j'\in\mathcal{J}}(\mathbb{E}(\Upsilon^{R}_{j'})) \ge -\epsilon \\
            0,& \text{if } \mathbb{E}(\Upsilon^{R}_{j_{a}}) - \max_{j'\in\mathcal{J}}(\mathbb{E}(\Upsilon^{R}_{j'})) \le -2*\epsilon \\
            1, & \text{otherwise}
	\end{cases}\\
	\label{eq:rew3}
	& \mathcal{V}^{L}_{j_{a}} =
	\begin{cases} 
        \mathcal{P}^{4}, & \text{if } \mathbb{E}(v_{j_{m}}) = 0 \\ 
        \mathcal{P}^{3},& \text{if } \mathbb{E}(v_{j_{r}}) = 0 \\
        \mathcal{P}^{2},& \text{if } \exists j'\in(\mathcal{J}/(j_{r}\cup j_{a}))(\mathbb{E}(v_{j'})) = 0 \\
        \mathcal{P}^{1}, & \text{otherwise} 
	\end{cases}
	\end{flalign}
	\par The main goal of the MDP model should be compatible with the objective function (Eq.~\ref{eq:objp1}) in order to obtain the maximization problem in P1. However, an MDP model may only attain that goal as a summation of the immediate rewards, $\mathbb{R}$, which can be found by following the policy $\Pi$ in each state-action pair ($\mathbb{S}x\mathbb{A}$). Therefore, we transform the objective function into Eq.~\ref{eq:rew1} which evaluates the recent action ($\mathbb{A}=j_{a}$) regarding the KPIs in the objective function.   
	\par P1 has three objectives, therefore the immediate reward should also consist of these three objectives. The first objective, increasing hovering time, is described as the total operating time before any of the ABSs have to recharge their battery, and in the problem definition section, we formulate that KPI as a MaxMin problem. We address that KPI with the battery reward levels (Eq. ~\ref{eq:rew2}) in the MDP model. That value has three levels [0,1,2] according to the expected remaining energy difference between the selected ABS $\mathbb{E}(\Upsilon^{R}_{j_{a}})$ and the ABS, which has the maximum battery energy $\max_{j'\in\mathcal{J}}(\mathbb{E}(\Upsilon^{R}_{j'}))$. If that energy difference is lower than a certain level ($\epsilon$), the reward function returns positive feedback. Therefore we can balance the battery energies between the ABSs. Lastly, hysteresis [-$\epsilon$,-2$\epsilon$] is introduced to prevent a ping-pong effect. 
	\par The second objective is minimizing mean delay, which is calculated for each action separately in the immediate reward calculation ($\Delta_{j_{a}}$). Therefore, we will get the total delay for all actions due to the cumulative reward calculation at the end of a simulation.
	\par The third objective depends on the expected deadline violation $\mathbb{E}(v_{j_{a}})$. It is equal to zero if the action ${j_{a}}$ is expected not to cause a deadline violation, and the reward function returns one for this objective. Otherwise, the reward function returns a negative value which is defined in Eq.~\ref{eq:rew3}. That value has four severity levels. The highest severity level, $\mathcal{P}^{4}$, occurs when we could have avoided the violation if we had chosen a MEC as a computational resource ($\mathbb{E}(v_{j_{m}}) = 1$). This case has the highest severity level because MEC devices have more powerful resources, and they don't have to consume valuable battery energy in order to process a task. The second-highest severity level, $\mathcal{P}^{3}$, occurs when a local resource could have processed that task without causing a violation. This case should have a higher penalty because offloading a task caused extra overhead and delay without benefiting from this action. The third level denotes the case if we have another ABS that could have processed that task without causing a violation. The final and the lowest severity level is a case where we expect a deadline violation as a result of any action.          
\end{itemize}
\subsection{Risk-Sensitive Learning Approach}
\par P2 has a deadline violation constraint (Eq.~\ref{eq:deadlineconstraint}) which can be achieved by only choosing an appropriate offloading action in each state-action pair. Therefore, that constraint is isolated from the other KPIs in order to limit the number of deadline violations to be lower than a threshold ($\mathbb{V}$). This separation can be addressed by a finite-horizon constrained MDP (CMDP) \cite{altman2004}. The CMDP is defined by tuple $\mathcal{F} = \{\mathbb{P}, \mathbb{A}, \mathbb{R}, \mathbb{C}, \mathbb{S}, \Pi \}$. The details of that tuple are:
\begin{itemize}
    \item \textbf{State Transitions:} $\mathbb{P}:\mathbb{S}_{1}x\mathbb{A}x\mathbb{S}_{2}\implies \mathbb{R}$
    \par State transitions are the same as in the Q-Learning approach that is explained in Section~\ref{sec:ql}.
    \item \textbf{State:} $\mathbb{S}= \{k, \{\Delta_{j'} | j'\in\mathcal{J^{+}} \}, \{\Upsilon^{L}_{j} | j\in\mathcal{J}\}, j_{a} \}$
    \par In addition to the states in the Q-Learning approach, the risk-sensitive method includes the recent offloading decision into the state space. Thus, we can split the states into two groups regarding risk expectation. Briefly, the ``risk states set" is a subset of the ``states set", and an agent falls into such a state when a deadline violation is expected ($\Phi=\{s \mid s \subset \mathbb{S} \textit{, and }\mathbb{E} (v_{j_{a}})\}$). Here, the expected deadline violation $(\mathbb{E} (v_{j_{a}}) = I(\alpha^{D}_{k} \le \Delta_{j_{a}}))$ is calculated by Eqs.(\ref{eq:delayviol1}-\ref{eq:delayviol2}).
    \item \textbf{Action:} $\mathbb{A}= \{j_{a} | j_{a} \in \mathcal{J^{+}}\}$
    \par The set of available resources ($j_{a}$) is the actions for the risk-sensitive approach.
    \item \textbf{Reward:} 
    \begin{flalign}
    \label{eq:rewnew}
    &\mathbb{R} = - \left[W * (\Upsilon^{L}_{j_{a}}-1) - \frac{1-W}{\Theta^{M}} * \Delta_{j_{a}} \right]
    \end{flalign}
	\par The main goal of the MDP model should be compatible with the objective function (Eq.~\ref{eq:objp2}) to obtain the maximization problem in P2. Therefore, we transform this objective function into Eq.~\ref{eq:rewnew} which evaluates the recent action ($\mathbb{A}=j_{a}$) in terms of the KPIs in the objective function. The details of this equation are explained in Section~\ref{sec:ql}. Another point we want to emphasize is that, unlike the Q-Learning approach, we reverse the reward function sign because the policy's exploitation phase chooses an action with the minimum q-value ($\argmin_{j_{a}\in \mathcal{J^{+}}}\mathbb{S}_{1}$).
	\item \textbf{Cost for Risk:} 
    \begin{flalign}
    \label{eq:riskcost}
    \mathbb{C} =
	\begin{cases} 
	        -\mathcal{V}^{L}_{j_{a}}, & \text{if } \mathbb{S}_{1} \in \Phi \\
            -1, & \text{otherwise}
	\end{cases}
    \end{flalign}
	\par In order to isolate and then prevent the risk in P2, we propose Eq.~\ref{eq:riskcost} to calculate the risk of each state-action pair. Calculation of the severity level ($\mathcal{V}^{L}_{j_{a}}$) is detailed in Section~\ref{sec:ql}.
    \item \textbf{Policy($\Pi$):} Since we have a separate cost function to evaluate the risk for each state-action pair, we have a q-table solely dedicated to the risk estimation. Fig.~\ref{fig:valiter2} shows the value iteration to calculate the risk in a smart farm. Meanwhile, we have another q-table to evaluate the other KPIs, shown in Fig.~\ref{fig:valiter3}. While these two isolated q-tables are used for the cumulative reward and risk calculations, we have to choose the same state-action pair in each step for these tables. Therefore we have a third q-table that picks the action for the corresponding state by using the classic epsilon-greedy approach (Fig.~\ref{fig:valiter4}). A value iteration process does not generate that q-table, instead, it is an aggregation of the previous two tables with a $\zeta$ weighting factor.
\end{itemize}
\begin{algorithm}
\caption{Dynamic update of $\zeta$ weighting value between the risk and reward}
\label{alg:zeta}
\begin{algorithmic}[1]
\STATE Given: $\zeta^{I}, \mathbb{V}, \lambda, N^{UP}, N^{EP}, \mathbb{G}$
\STATE Output: $Q^{D}, Q^{R}, Q^{C}$
\STATE $episode = 0, \zeta = \zeta^{I}$
\WHILE{$episode <N^{EP}$}
\STATE $Q^{D}$ :: value iteration for risk \qquad(Fig.~\ref{fig:valiter2})
\STATE $Q^{R}$ :: value iteration for reward \quad(Fig.~\ref{fig:valiter3})
\STATE $Q^{C} = \zeta * Q^{D} + (1- \zeta) * Q^{R}$\qquad(Fig.~\ref{fig:valiter4})
\STATE $episode = episode + 1$
\IF {$episode \equiv 0 \pmod{N^{UP}}$}
\IF {$\sum\limits_{\taskset}v_{\task } + \mathbb{G} \leq \mathbb{V}$}
\STATE $\zeta = \min(\zeta - \lambda, 0)$
\ELSE
\STATE $\zeta = \max(\zeta + \lambda, 1)$
\ENDIF
\ENDIF
\ENDWHILE
\end{algorithmic}
\end{algorithm}
\par Choosing a decent $\zeta$ is essential to improve the convergence of the proposed risk-sensitive solution. Deciding on a higher $\zeta$ prioritizes the risk side of the problem, while choosing a lower one may improve the other KPIs. We decide on a proper $\zeta$ with a dynamic approach shown in Algorithm~\ref{alg:zeta}. Given parameters $\zeta^{I}, \mathbb{V}, \lambda, N^{UP}, N^{EP}, \mathbb{G}$ are the initial weighting value, deadline violation threshold, updating step size, updating frequency, the number of episodes, and the gap value, respectively.  
\par This algorithm starts with the initialization of episode and $\zeta$ values. Then, for each episode, we calculate the three q-tables as they are illustrated in Figs.~(\ref{fig:valiter2}, \ref{fig:valiter3}, \ref{fig:valiter4}). Line~9 checks the updating frequency, and then Line~10 compares the number of deadline violations with the threshold. In that equation, we add an additional gap ($\mathbb{G}$) due to the uncertainty property of the tasks. Therefore, the algorithm guarantees Eq.~\ref{eq:deadlineconstraint} in some cases where the IoT devices demand to process an extremely high number of tasks. Following this comparison, Line~11 decreases the weight of the risk in the case of where Eq.~\ref{eq:deadlineconstraint} is guarenteed. Otherwise, Line~13 increases the weight of the risk to attain that constraint.          
\par The proposed risk-sensitive method is an episodic model-free reinforcement learning algorithm. Jin et al. \cite{Jin2018} prove that the worst-case complexity of this algorithm is $\tilde{\mathcal{O}}(H^{2}\sqrt{|\mathbb{A}| |\mathbb{S}| N^{EP}})$, where $H$ is the number of steps in an episode, $|\mathbb{A}|$, and $|\mathbb{S}|$ are the action and state space sizes, and $N^{EP}$ is the number of episodes. We train our risk-sensitive method offline with a collected dataset from the simulation platform. Each of these datasets has a number of tasks defining the number of steps ($H$) in an episode. Thus, our algorithm's complexity quadratically increases with the number of tasks in each episode. According to Section IV.B., with the increasing number of ABSs ($J$), the state space ($|\mathbb{S}|$) changes quadratically, and the action space ($|\mathbb{A}|$) changes linearly. Therefore, computation time grows as $J^{\sqrt{3}}$ with the number of ABSs in the problem. Meanwhile, the number of episodes ($N^{EP}$) is determined according to the algorithm's convergence, which is not only related to the size of the given data. The convergence of the algorithm (training phase) is also affected by the content of the data. Therefore, we empirically determine the value of $N^{EP}$.
\section{Numerical Results}
\par \subsection{Simulation Platform}
We chose Omnet++ \cite{omnetpp} as a discrete event simulator to simulate the IoT traffic. In addition, we used Simu5G libraries \cite{Nardini2020}, which is developed over Omnet++, to simulate a 5G channel model between IoT devices, ABSs and MEC platform. These libraries implement the TR 36.873 specification \cite{3GPP2018} to calculate the channel fading in new radio numerology ($\mu=0$) of a 5G physical layer. Shadow fading standard deviation and carrier frequency are chosen at $4 dB$ and $2 GHz$ for line-of-sight communication, respectively.
Table~\ref{tab:SimParams} presents the parameters used in the simulations. In particular, we looked at a smart farm scenario with four ABSs ($J$=4) and one MEC ($L$=1) devices. We focused on three types of image processing operations ($\mathcal{K}$) that have different arrival rates, deadlines and processing times. Interarrival times between the tasks are exponentially distributed with a mean of $(1/\lambda)$. Deadlines and processing times are constants. The results are generated as the average of ten runs with different seeds and we set the simulation time to be 50 seconds ($T=50s$). 

\begin{table}
    \centering
    \caption{\label{tab:SimParams} Simulation parameters.}
    \begin{tabular}{l|c|c|c|c}
    Task Type & $(1/\lambda)$ & $\alpha^{D}_{\task}$ & \begin{tabular}{@{}c@{}}$\alpha^{P}_{\task}$ \\ (ABS)\end{tabular} & \begin{tabular}{@{}c@{}}$\alpha^{P}_{\task}$ \\ (MEC)\end{tabular}\\
    \hline
    Fire detection &  0.25s & 1s & 0.1s & 0.05s\\
    Pesticide detection &  0.25s & 2s & 0.2s & 0.1s \\
    Growth monitoring & 0.5s & 15s & 1.5s & 0.75s \\
    \end{tabular}
\end{table}

\begin{table}
    \centering
    \caption{\label{tab:EngParam} Energy consumption parameters.}
    \begin{tabular}{c|c|c|c|c|c}
    ABS No ($j'$)&$\Upsilon^{B}_{j'}$&$\Upsilon^{H}_{j'}$&$\Upsilon^{A}_{j'}$&$\Upsilon^{I}_{j'}$&$\Upsilon^{C}_{j'}$\\
    \hline
    0&570&211&17&4320&12960\\
    1&570&211&17&4320&12960\\
    2&627&211&17&4320&12960\\
    3&627&211&17&4320&12960\\
    \end{tabular}
\end{table}  
\subsection{Baselines}
\subsubsection{Round Robin}
Every computational resource located at the ABSs and the MEC is utilized in a round robin fashion when assigning the tasks.


\subsubsection{Lowest Queue Time and Highest Energy First}
This algorithm considers the destinations' remaining energy level and queue time. As stated in \cite{Nguyen2021}, queue time is the total time it takes for the resource to compute all of the tasks in its queue. Every ABS stores the remaining energy level and queing time of every possible offloading destination. The ABSs regularly update their neighbours with such information. The decision making algorithm begins by finding the lowest queue time amongst the neighbouring resources. If an ABS's queue time is lower than the current ABS's queue time by at least 0.5s, we will use the this value as the lowest queue time. Else, the algorithm will use the current ABS's queue time as the lowest queue time. Next, the algorithm will find the neighbour that has the highest remaining energy level and the queue time is equal or less than the lowest queue time. If such a node exists and its energy level is higher than the current ABS's energy level by at least 1\%, then the current ABS will offload the task to that node. If not, then the task will be computed locally.

\subsubsection{Energy-Centric Approach}
This method prioritizes the Max-min fairness over the deadline violation minimization. Therefore, it provides a higher hovering time for the smart farm. Implementation is based on the proposed risk-sensitive method. Meanwhile, the size of the difference between the highest energy battery and the lowest energy battery is defined as risk. Thus Algorithm~\ref{alg:zeta} keeps the $\zeta$ value at higher levels while this gap is larger than a certain threshold.
 
\subsection{Energy Consumption Parameters}
We used the parameters found in Table~\ref{tab:EngParam} and Eq.~\ref{eq:encalc} in order to model an ABS's remaining battery energy. By considering limited simulation time, idle and busy CPU energy consumptions are defined as ABSs that would be run for ten hours to expose the performance differences between the techniques.  

\begin{figure*}
  \begin{subfigure}{.32\textwidth}
  \includegraphics[width=1.0\textwidth]{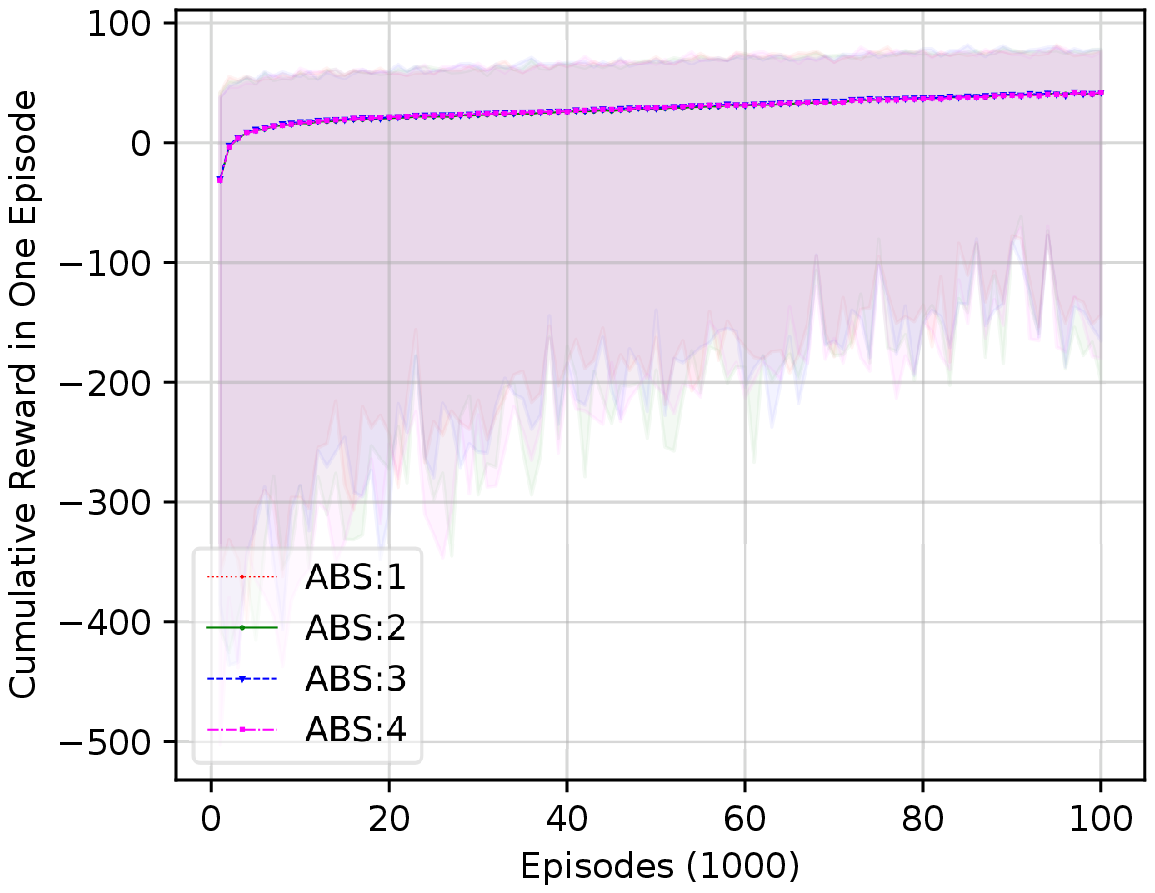}
  \caption{Q-learning convergence.}\label{fig:conv1}
  \end{subfigure}
  \begin{subfigure}{.32\textwidth}
  \includegraphics[width=1.0\textwidth]{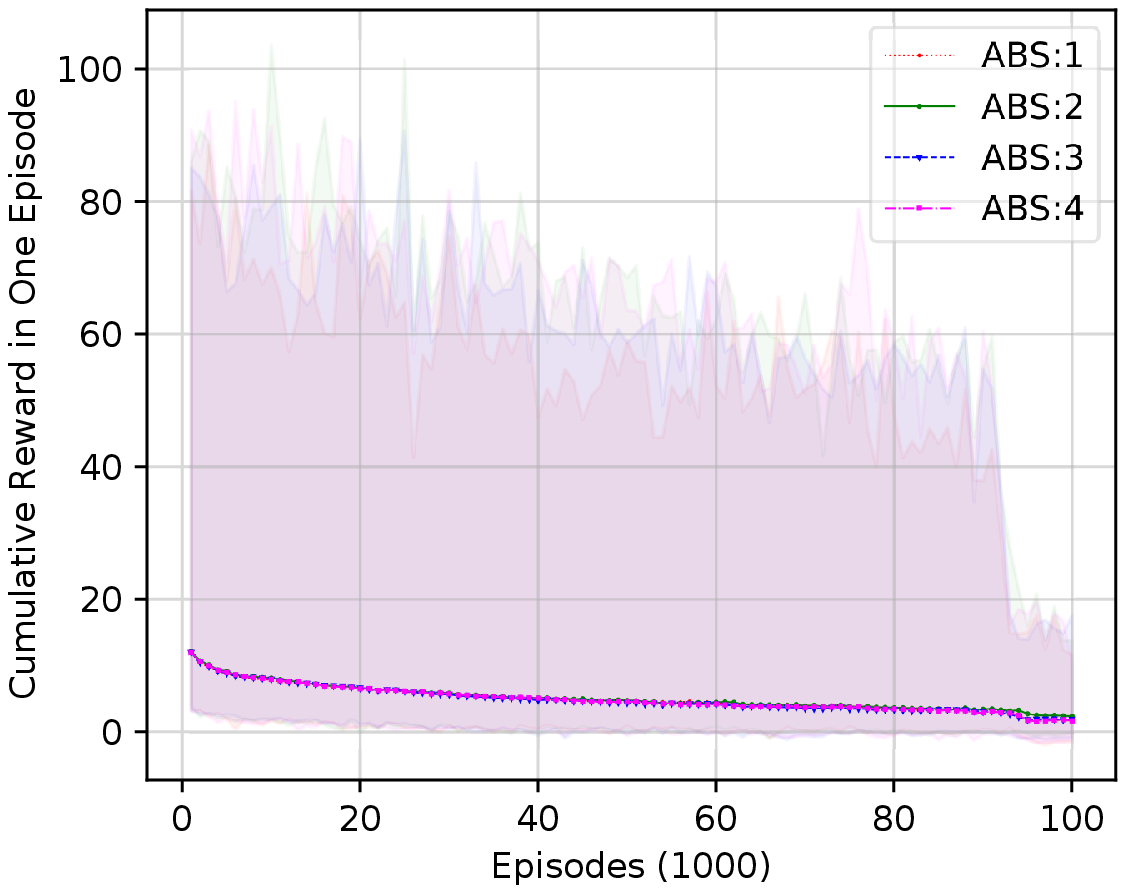}
  \caption{Reward convergence in Risk-sensitive.}\label{fig:conv2}
  \end{subfigure}
  \begin{subfigure}{.32\textwidth}
  \includegraphics[width=1.0\textwidth]{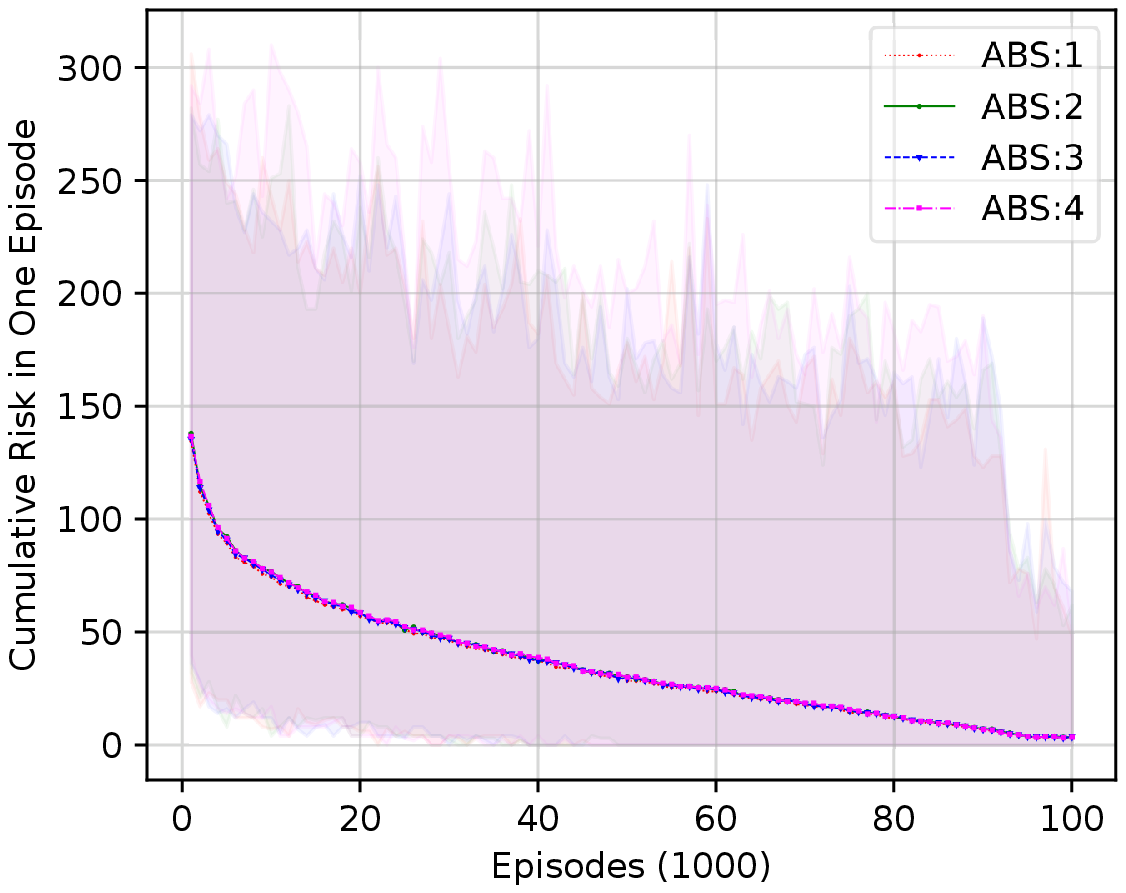}
  \caption{Risk convergence in Risk-sensitive.}\label{fig:conv3}
  \end{subfigure}
  \caption{Convergences of Q-learning and Risk-sensitive method. }
  \label{fig:conv}
\end{figure*}

\begin{figure*}
  \begin{subfigure}{.48\textwidth}
  \includegraphics[width=0.9\textwidth]{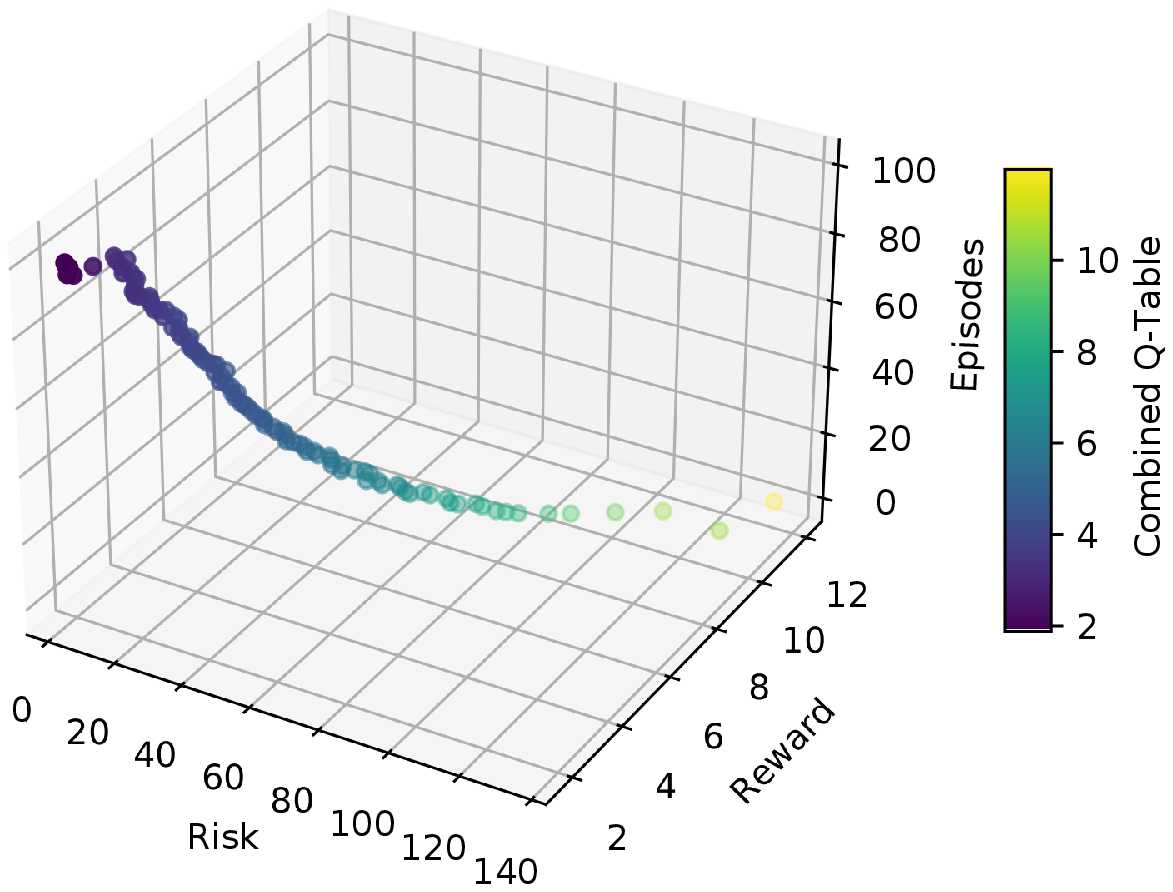}
  \caption{Risk-sensitive method.}\label{fig:convdyn1}
  \end{subfigure}
  \begin{subfigure}{.48\textwidth}
  \includegraphics[width=0.9\textwidth]{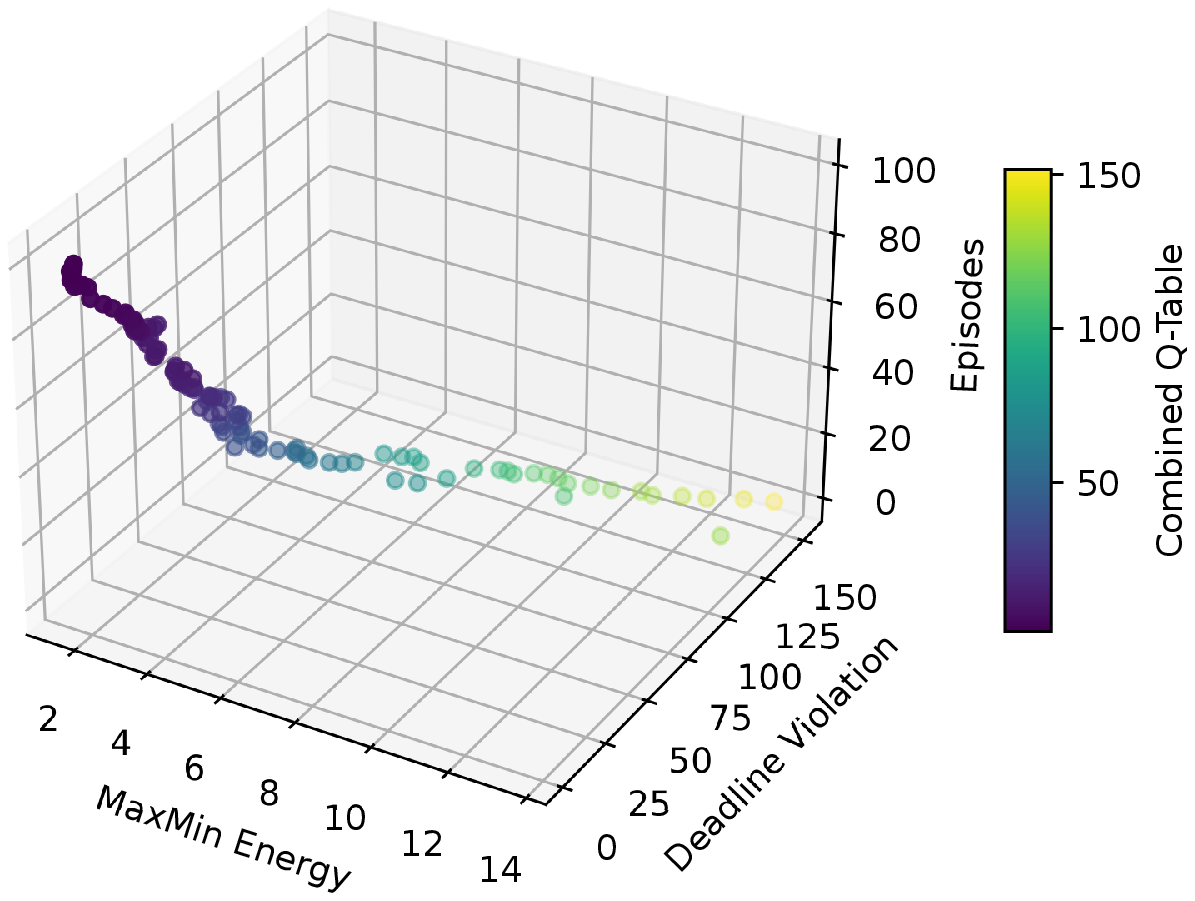}
  \caption{Energy-centric method.}\label{fig:convdyn2}
  \end{subfigure}
  \caption{Dynamically change of convergences between reward and risk q-tables. }
  \label{fig:convdyn}
\end{figure*}

\subsection{Convergence of Machine Learning Approaches} 
\begin{table}
    \centering
    \caption{\label{tab:mlParams} Machine Learning Parameters.}
    \begin{tabular}{c|c|c}
    Parameter & Notation & Value \\ 
    \hline
    Learning Rate& $\alpha$ & $0.05$\\
    Discount & $\gamma$ & $0.85$\\
    Number of Episodes & $N^{EP}$ & $100k$\\
    Updating frequency & $N^{UP}$ & $1k$\\
    Initial weighting value & $\zeta^{I}$ & $0.00$\\
    Step Size & $\lambda$ & $0.02$\\
    Max. Deadline Violation & $\mathbb{V}$ & $3$\\
    Gap value & $\mathbb{G}$ & $2$\\
    Weighting value for Q-Learning & $W$ & $0.5$\\
    Mean delay $\&$ deadline violation scalers& $\Theta^{M}, \Theta^{D}$ & $1,1$\\
    Deadline violation severity levels& $\mathcal{P}^{4,3,2,1}$& $-\{4,3,2,1\}$\\
    \end{tabular}
\end{table}  
\par We use an offline learning approach for machine learning (ML) solutions. First, we generate one hundred different data sets with Omnet++ for the learning phase. Then, we run Q-Learning, Energy-centric and Risk-sensitive methods for 100k episodes, in which each episode uses another data set to prevent an overfitting problem. Then, we transfer the generated tables in this learning phase to the test phase. Finally, in the test phase, we use Omnet++ to simulate a real 5G wireless channel model and evaluate the generated tables according to several KPIs. That KPI evaluation is detailed in the following subsection. Table~\ref{tab:mlParams} represents the numerical values used in ML algorithms.
\par Fig.~\ref{fig:conv1} illustrates the convergence of the Q-learning baseline method. As seen, this method practices the $\argmax$ approach to find the state-action value. If we turn to the risk-sensitive method, Fig.~\ref{fig:conv2} and Fig.~\ref{fig:conv3} show the convergence of reward and risk tables in the learning phase, respectively. Owing to multi-actor-based approach and use of different datasets in each episode, the variations in the cumulative reward and risk represent the extreme levels in both tables. On the other hand, it is seen that each actor (in other terms, each ABS) learns at the same speed and approaches the zero level.
\par Fig.~\ref{fig:convdyn1} shows the trend of convergences in these two tables. As mentioned in Algorithm~\ref{alg:zeta}, we prioritize the risk table at the beginning of the learning phase. Fig.~\ref{fig:convdyn1} confirms that algorithm; while the cumulative reward reduces in the risk table 57\% (from 140 to 60), it is only 20\% (from 14 to 9) in the reward table at the same episode. We should note that the tradeoff between different KPIs may adaptively change with Algorithm~\ref{alg:zeta} for different smart farm scenarios. This can be seen easily in Fig.~\ref{fig:convdyn2}. This figure shows the learning trend of the Energy-centric approach, which prioritizes the MaxMin fairness of energy usage in the batteries of ABSs. 

\subsection{KPI Evaluations} 
\par
\begin{figure}
\centering
\includegraphics[width=0.48\textwidth]{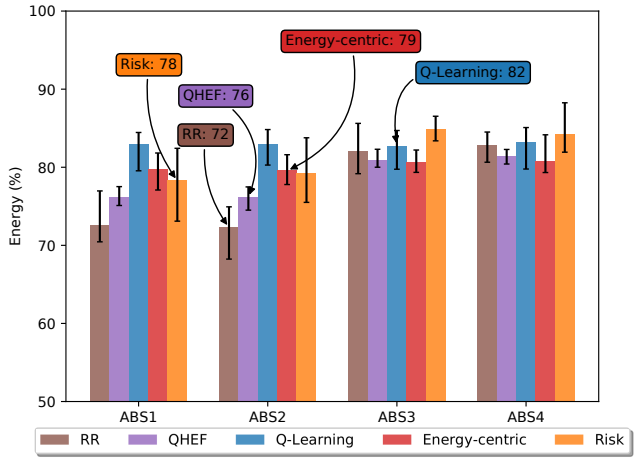}
\caption{\label{fig:hovtime} Remaining energy levels in different methods.}
\end{figure}

 Fig.~\ref{fig:hovtime} shows the remaining energy levels in the batteries of ABSs for different solution methods. As it is explained in the problem definition, we aim to increase the hovering time of ABSs. Then, we propose a MaxMin solution in which we maximize the energy of the ABS, which has the minimum battery. That value is 78\% for our proposed risk-sensitive method, which is higher than the heuristics but lower than the Q-Learning approach. The main reason for that output is prioritizing deadline violations in risk-sensitive solutions. Also, the energy-centric baseline performs better than the risk-sensitive approach and provides an extensive fairness between the ABSs. Meanwhile, the risk definition of that approach might be revised for outperforming the classical Q-learning method.
\par
\begin{figure}
\centering
\includegraphics[width=0.48\textwidth]{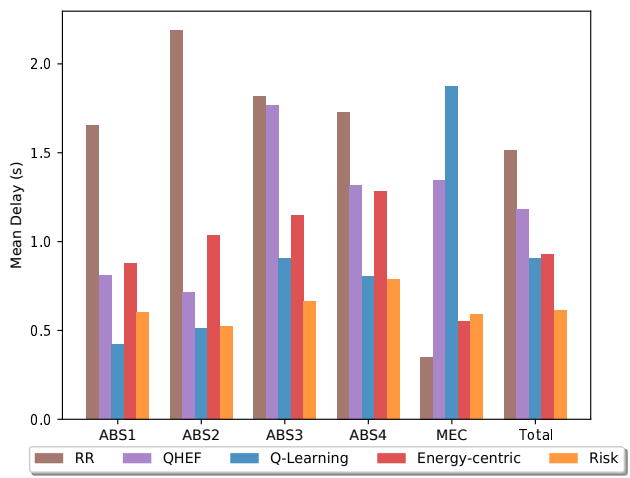}
\caption{\label{fig:meandelay} Mean delay in different methods.}
\end{figure}

 Fig.~\ref{fig:meandelay} illustrates the mean delay KPI. It is clear that ML approaches outperform the heuristics. Meanwhile, there is no decisive optimal ML solution in terms of mean delay minimization in different computational resources. Q-Learning has better performance in the ABSs, but it has a dramatically higher delay in MEC. This is because the Q-Learning policy sends more tasks to MEC. On the other hand, the risk-sensitive approach provides lower overall delay than all other methods.
\par
\begin{figure}
\centering
\includegraphics[width=0.48\textwidth]{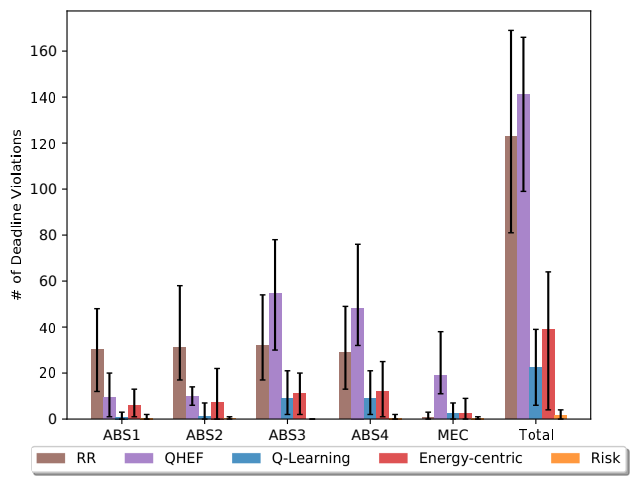}
\caption{\label{fig:delay} Number of deadline violations.}
\end{figure}

 The main performance improvement of the risk-sensitive approach can be seen in Fig.~\ref{fig:delay}. This figure demonstrates the deadline violations in different ABSs and MEC. Although the Q-Learning and Energy-centric methods outperform the heuristics, they can not guarantee that the number of deadline violations will be lower than a threshold. In contrast, this figure demonstrates that our risk-sensitive solution establishes risk-free offline learning for a multi-objective optimization problem.

\subsection{Upper Bound Comparison} 
\begin{table}
    \centering
    \caption{\label{tab:ml} Upper Bound Comparison (P1: Eq. \ref{eq:objp1}, P2: Eq. \ref{eq:objp2}, M: MILP Solver, R: Risk-Sensitive ML, Q: Q-Learning ML)}
    \begin{tabular}{ c|c|c||c|c|c } 
	 Problem & Method & $W$ & $\min\limits_{ j' \in J} \Upsilon^{R}_{j'}$ &$\delta$&$\sum\limits_{\substack{j\in\mathcal{J}\\ t\in\mathcal{T}}}v_{jt}$\\
	 \hline
	 P1 & M & 1 &  $98.2\%$   & 1.89s   &$72$\\
	 P1 & Q & 1 &  $98.2\%$   & 4.01s   &$89$\\ 
	 P1 & M & 0.5 & $97.0\%$ & 0.92s &$14$\\
	 P1 & Q & 0.5 & $97.7\%$ & 0.68s &$42$\\
	 P1 & M & 0 &  $96.3\%$  & 0.84s &$9$\\
	 P1 & Q & 0 &  $97.0\%$  & 0.44s &$26$\\ \hline
	 P2 & M & 1 &  $97.3\%$   & 0.88s   &$15$\\
	 P2 & M & 0.5 & $96.7\%$ & 0.89s &$15$\\
	 P2 & M & 0 & $96.3\%$   & 0.85s &$15$\\
	 P2 & R & N &  $96.5\%$ & 0.45s   &$13$\\ 
	 \hline
    \end{tabular}
\end{table}

We chose Gurobi Solver to find the upper bounds for the maximization problems \cite{GurobiOptimization2021}. Owing to the fact that these problems are NP-hard, a MILP solver may provide a solution for small space problems \cite{Pamuklu2020}. Therefore, we had to limit the simulation time to four seconds ($T=4s$), and we ran the solver for 48 hours for each case to find the optimum solution. Furthermore, in order to increase the number of tasks in this shorter time duration, we chose interarrival time as 0.125s for all task types. Finally, we reduce the deadlines for fire detection ($0.2s$) and pesticide detection tasks ($0.6s$) to make the problem harder than the standard configuration.
\par The top part of Table~\ref{tab:ml} shows the upper bounds of P1 (Eq.\ref{eq:objp1}) with different weight factors ($W$). The first row ($W=1$) represents the maximum hovering time for P1. Our proposed ML solution for P1 can reach that level (second row) if we only add the hovering time maximization in the reward function. On the other hand, in a balanced solution ($W=0.5$), it is seen that the ML solution tends to favor hovering time. Although we can get higher hovering time by ML, the MILP solution outperforms this method in terms of mean delay and deadline violation minimization. Lastly, if we remove the hovering time KPI from the multi-objective function ($W=0$), the MILP solver reduces the mean delay and the number of violations significantly.   
\par The bottom section of Table~\ref{tab:ml} shows the upper bounds of P2 (Eq.\ref{eq:objp2}). First, we find a feasible solution with a significantly higher deadline violation threshold ($\mathbb{V}=15$) than the solution of MILP P1 \footnote{P2 has an additional constraint for limiting the number of deadline violations. Adding an extra constraint usually increases the solution space of a MILP solution. Therefore it may need more time to find a feasible solution. It has to be reminded that we limit the MILP solutions by 48 hours.}. It is undoubtedly seen that the risk-sensitive ML surpasses the MILP solution regarding minimizing the number of deadline violations. Moreover, the ML solutions are far better for reducing the mean delay. In addition, its hovering time maximization performance is almost the same as the MILP solution.     
\subsection{Processing Time and Deadline Alterations} 
\begin{figure}
\centering
\includegraphics[width=0.48\textwidth]{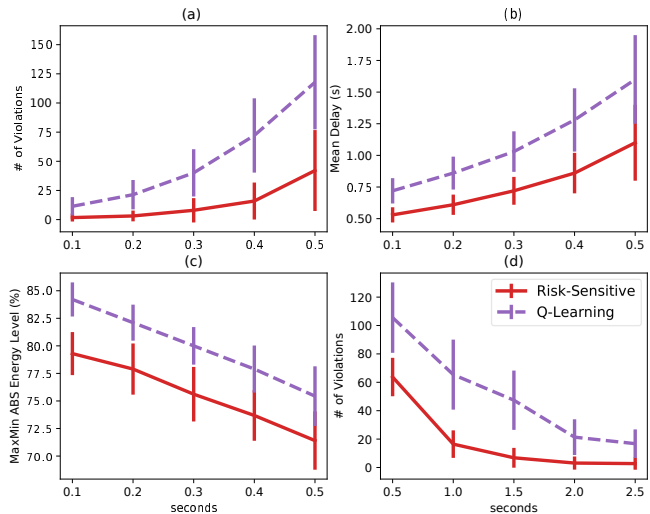}
\caption{\label{fig:eval} Impact of pesticide detection task parameter variations. (a), (b), (c): Increasing processing time. (d): Increasing deadline.}
\end{figure}

It has been claimed that training an ML solution with offline data provides an efficient and reliable solution for a wireless network problem \cite{O-RANWG22021,Bonati2021}. Therefore, we trained our ML solutions with preliminary data and then solve those finding solutions within a simulation platform. In this subsection, we provide a kind of stress test for our ML solutions. We generate a one-hundred offline dataset that has different processing time and deadline for pesticide detection (PD) task (Fig~\ref{fig:eval}). The results shows that the risk-sensitive solution adapts better than Q-Learning in most KPIs for the changes of these parameters. 

\section{Conclusion}
This study focused on a smart agriculture scenario that uses ABSs to augment the computational resources of IoT devices that need to process deadline-critical image processing tasks. We defined a problem definition that made sure that the tasks are completed before their deadlines and improved the hovering time of these ABSs. We addressed this using a CMDP, and then proposed a risk-sensitive approach for this CMDP that constrained the number of deadline violations. The results show that a risk-sensitive approach is feasible for guaranteeing a constraint-based KPI while providing an energy-efficient solution for the ABSs. As future work, we are planning to provide a solution for more extensive state-space problems.

\section*{Acknowledgement}
This  work  is  supported  by  MITACS Canada Accelerate program under collaboration with Nokia Bell Labs. 

\bibliography{main}

\end{document}